\documentclass{cs23proc}

\usepackage{kantlipsum}
\usepackage{url}
\usepackage{hyperref}
\usepackage{orcidlink}
\editors{Takeru Suzuki and the Cool Stars 23 Organizing Team}
\publisher{Zenodo}
\conference{The $23^{\mathrm{rd}}$ Cambridge Workshop on Cool Stars, Stellar Systems, and the Sun (Cool Stars 23)}
\conferencedate{2026}

\title{Open Clusters as Laboratories for Cool Star Evolution: \\
Highlights from the Cool Stars 23 Splinter Session}
\author{Deepak~Chahal$^{1}$ \orcidlink{0000-0002-3612-3622}, 
Beatrice~Caccherano$^{1}$ \orcidlink{0009-0001-3865-0119}, 
Edward~Gillen$^{1}$ \orcidlink{0000-0003-2851-3070}, 
Dario~J.~Fritzewski$^{2}$ \orcidlink{0000-0002-2275-3877}, Robin~D.~Jeffries$^{3}$ \orcidlink{0000-0001-5668-1288},
Kevin~R.~Covey$^{4}$ \orcidlink{0000-0001-6914-7797}, 
Andrew~Boyle$^{5}$ \orcidlink{0000-0001-6037-2971},
Lyra~Cao$^{6}$ \orcidlink{0000-0002-8849-9816},
Federica~Chiti$^{7}$ \orcidlink{0000-0001-5180-2271},
Alexander~Hughes$^{1}$ \orcidlink{0009-0001-3561-5035},
Leslie~Moranta$^{8}$ \orcidlink{0000-0001-7171-5538}, 
Natalie~R.~Myers$^{9,10}$ \orcidlink{0000-0001-9738-4829},  
Emily~K.~Pass$^{11}$ \orcidlink{0000-0002-1533-9029},
Phil~Van-Lane$^{12,13}$ \orcidlink{0009-0009-4567-9946},
Fábio~C.~Wanderley$^{14}$ \orcidlink{0000-0003-0697-2209},
Mackenna~L.~Wood$^{15}$ \orcidlink{0000-0001-7336-7725},
Stephanie~T.~Douglas$^{16}$ \orcidlink{0000-0001-7371-2832},
David~Montes$^{17}$ \orcidlink{0000-0002-7779-238X},
Loredana~Prisinzano$^{18}$ \orcidlink{0000-0002-8893-2210}
}

\affiliation{$^{1}$Astronomy Unit, Queen Mary University of London, Mile End Road, London E1 4NS, UK\\
$^{2}$Institute of Astronomy, KU Leuven, Celestijnenlaan 200D, 3001, Leuven, Belgium\\
$^{3}$Astrophysics Group, Keele University, Keele, Staffordshire ST5 5BG\\
$^{4}$Department of Physics \& Astronomy, Western Washington University, 516 High St., Bellingham WA, 98225, USA\\
$^{5}$Department of Physics and Astronomy, The University of North Carolina at Chapel Hill, Chapel Hill, NC 27599, USA\\
$^{6}$Department of Physics and Astronomy, Vanderbilt University, Nashville, TN 37235, USA\\
$^{7}$Institute for Astronomy, University of Hawai‘i at Mānoa, 2680 Woodlawn Drive, Honolulu, HI 96822, USA\\
$^{8}$Planétarium Rio Tinto Alcan, Espace pour la Vie, 4801 av. Pierre-de Coubertin, Montréal, Québec, Canada\\
$^{9}$William H.\ Miller III Department of Physics \& Astronomy,
Johns Hopkins University, 3400 N Charles St, Baltimore, MD 21218, USA. \\
$^{10}$Department of Physics and Astronomy, Texas Christian University, TCU Box 298840 
Fort Worth, TX 76129, USA\\
$^{11}$Kavli Institute for Astrophysics and Space Research, Massachusetts Institute of Technology, Cambridge, MA 02139, USA\\
$^{12}$David A. Dunlap Department of Astronomy \& Astrophysics, University of Toronto, 50 St. George St, Toronto, ON M5S 3H4, Canada\\
$^{13}$Department of Astronomy \& Astrophysics, University of California San Diego, 3183 Matthews Ln, La Jolla, CA 92093, USA\\
$^{14}$Universidade Federal do Rio de Janeiro, Observat\`orio do Valongo, Brazil\\
$^{15}$Museum of Science \& Industry, Tampa, FL 33617, USA\\
$^{16}$Department of Physics, Lafayette College, Easton, PA 18042, USA\\
$^{17}$ Departamento de Física de la Tierra y Astrofísica—Facultad de Ciencias Físicas, Universidad Complutense de Madrid, E-28040, Madrid, Spain\\
$^{18}$INAF-Osservatorio Astronomico di Palermo, Piazza del Parlamento 1, 90134, Palermo, Italy\\
}

\shorttitle{Open Clusters as Laboratories for Cool Star Evolution}
\shortauthors{Chahal et al.}

\abs{ 
 
Open clusters remain among the most powerful laboratories for calibrating how fundamental stellar properties such as rotation, magnetic activity, surface chemistry, and internal structure evolve over a star's pre-main-sequence to main-sequence and post-main-sequence lifetime. Because cluster members share a common age, galactic environment, and initial composition, they provide empirical anchors upon which age-dating techniques such as gyrochronology, activity-age relations, and chemical clocks are built. This paper summarizes the Cool Stars 23 splinter session ``Open Clusters as Laboratories for Cool Star Evolution'', held on 15 June 2026 in Tokyo, Japan. The session comprised one invited review, ten contributed talks, and ten poster pop-up presentations, organized around three themes: the evolution of rotation, magnetic activity, and chemical abundances (Li-depletion, [C/N], [Y/Mg], and neutron-capture elements). We close with a summary of the open questions identified in the panel discussion and the observational and theoretical work that new surveys (e.g. \textit{Gaia} DR4, 4MOST, WEAVE) and missions (e.g. PLATO, Roman) will enable over the coming years. Together, these efforts will help assess the current state of the field and shape its future directions.}
\begin{document}
\maketitle
\section{Introduction} 
Galactic open clusters are groups of stars that formed from the same giant molecular cloud and therefore share approximately the same age, distance, and initial chemical composition. Their stars are loosely bound by mutual gravitational attraction, and while most open clusters survive for only a few hundred million years, the most massive systems can persist for several billion years (\citep{Vandenberg2004}, with the most notable examples being M 67 and NGC 188). Because they contain stars spanning a wide range of masses but with broadly common ages and metallicities, open clusters provide natural laboratories for studying stellar evolution.

Determining precise ages for cool stars remains one of the longstanding challenges in stellar astrophysics. This limitation also hinders our understanding of how key stellar properties, such as rotation, magnetic activity, and surface chemical abundances, evolve over time. Unlike fundamental properties such as effective temperature, radius, or composition, age cannot be measured directly for main-sequence stars. Instead, it must be inferred from secondary, time-dependent observables that change over a star’s lifetime and require empirical calibration using stellar populations of independently known age \citep[see][for a review]{Soderblom2010}. Open clusters, with their approximately coeval populations, provide critical benchmarks for this calibration. Comparing clusters spanning a broad range of ages allows us to trace the evolution of stellar properties such as rotation, magnetic activity, and surface chemical abundances, while simultaneously testing stellar models and age-dating techniques.

Recent observations and characterization of large samples of open clusters \citep[e.g. ][]{Cantat-Gaudin2018, Kounkel2019, Cantat-Gaudin2020a, Cantat-Gaudin2020b, Hunt2023, Hunt2024} spanning ages from a few million years to several billion years provide a unique opportunity to investigate the evolution of stellar properties across different stages of stellar evolution. Among the many properties that can be explored, three show measurable and systematic evolution across the open-cluster age sequence, forming three central themes of this splinter session. 

\emph{Stellar Rotation}: Stellar rotation evolves throughout the pre-main, main, and post-main sequence lifetimes as stars contract and angular momentum is redistributed within the stellar interior and lost through magnetized stellar winds. Time-series observations from space-based missions such as CoRoT \citep[e.g.][]{Affer2012}, Kepler \citep[e.g.][]{McQuillan2014}, and TESS \citep[e.g.][]{tars_Aboyle}, together with extensive ground-based surveys including ASAS-SN \citep[e.g.][]{Jayasinghe2020} and ZTF \citep[e.g.][]{Chen2020, Chahal2022}, have enabled the measurement of rotation periods for large samples of cool stars. Over the past decade, substantial efforts have been made to detect and characterize rotation in open-cluster populations (\citealt{Rebull2016}; \citealt{Douglas2017}; \citealt{Fritzewski2021}; \citealt{Curtis2020}; NGTS Cluster survey - \citep{Gillen2020, Smith2023, Hughes2026}; MOCA-DB - \citealt{Gagne2026}), providing crucial empirical benchmarks for studying rotational evolution. Open clusters have also been central to the development and calibration of gyrochronology models from their inception (\citealt{Skumanich1972}; \citealt{Barnes2003}; \citealt{Bouma23_emp_lim}; \citealt{VanLane2025}; \citealt{Gruner2026}). Recent studies based on cluster rotation sequences suggest that rotational spin-down depends strongly on stellar mass and age, with possible departures from the simple monotonic evolution predicted by classical gyrochronology \citep{Barnes2003, Barnes2007, Barnes2010}, including periods of stalled or weakened spin-down \citep{Curtis2019a, Curtis2020}. These observations highlight the limitations of simple gyrochronology prescriptions and point towards the need for more sophisticated models that incorporate processes such as mass-dependent magnetic braking, braking saturation, and core–envelope coupling (\citealt{Spada2015}).

\emph{Magnetic activity}: The interplay between stellar rotation and magnetic fields, and its role in driving magnetic activity, remains poorly understood. Magnetic activity manifests across stellar atmospheres through a range of phenomena, including starspots and faculae on the photosphere, chromospheric emission, and energetic events such as flares and coronal mass ejections (CMEs) originating in the corona (\citealt{Hall2008}; \citealt{deGrijs2021}). The physical mechanisms governing the generation and evolution of these activity features, and their dependence on stellar mass and evolutionary state, remain uncertain. Open clusters provide an ideal benchmark for investigating these processes, as their coeval stellar populations enable the evolution of magnetic activity to be studied as a function of stellar mass and age. In particular, rotational modulation caused by starspots and faculae in stellar light curves \citep{McQuillan2014}, combined with medium- to high-resolution spectroscopic surveys such as Gaia-ESO, GALAH, and LAMOST \citep{Frasca2025}, now enables the evolution of magnetic activity to be traced across a broad range of stellar masses and cluster ages.  Recent work has, for example, investigated whether the observed stall in rotational spin-down is associated with changes in magnetic activity \citep{Cao2023, Chahal2023, Gody-Rivera2026}, providing insights into the coupled evolution of stellar rotation and magnetism across stellar masses and ages. Several studies have investigated stellar magnetic activity in open clusters across different layers of the stellar atmosphere, from the photosphere to the corona \citep{Nunez2024, Fritzewski2025}. Open clusters provide a particularly valuable framework for such studies, as their coeval populations allow the evolution of magnetic activity to be examined alongside stellar rotational evolution. 

\emph{Chemical abundances}: Chemical abundances provide important insights into stellar evolution and Galactic archaeology. Surface lithium, which is readily depleted at relatively low temperatures, offers a sensitive probe of stellar mixing and evolutionary processes \citep{Jaffries23}. In giant stars, the C/N ratio can serve as an age proxy, as the first dredge-up along the giant branch is strongly mass-dependent \citep{Casali2020, Hayden2022, Taylor2022}, although its calibration requires precise asteroseismic ages. Similarly, studies of nearby stars have established age–[$\alpha$/Fe] relations, with [$\alpha$/Fe] providing a useful age proxy for $\alpha$-enhanced populations \citep{Bensby2014,Hayden2022}. However, these chemical age indicators remain only partially calibrated against precise ages. Open clusters, with their well-constrained ages and chemical homogeneity, therefore provide an ideal benchmark for calibrating and validating these relations.


 
 
The timing of this session was motivated by a wave of new and forthcoming data. Photometric surveys from \emph{TESS} and ground-based facilities (e.g., NGTS), together with legacy \emph{Kepler}/K2 data, and recent and ongoing spectroscopic surveys (Gaia-ESO, GALAH, LAMOST, APOGEE), already span open clusters from $\approx$1~Myr to $\approx$3~Gyr in age. Over the coming years, the large-scale 4MOST and WEAVE spectroscopic surveys will deliver high- and low-resolution spectra for tens of thousands of additional cluster members, PLATO will observe many tens of open clusters with long-baseline, high-precision time series observations, while \emph{Gaia} DR4 astrometry and photometry will sharpen cluster membership to unprecedented precision. Combined with advances in statistical and machine-learning membership and age-inference methods, these datasets are poised to produce the most detailed picture yet of how rotation, activity, and chemistry evolve together across stellar mass and age.
 
This splinter session brought together researchers working across these three themes to share recent results and identify open questions and priorities for the field. Section~\ref{sec:overview} gives an overview of the open clusters included in the invited review compilation; Sections \ref{sec:rotation}, \ref{sec:activity} $\&$ \ref{sec:abundances} present the contributed talks and poster highlights organized by theme; Section~\ref{sec:panel} summarizes the closing panel discussion; and Section~\ref{sec:summary} looks ahead to how upcoming surveys and missions will build on the work presented here.


\section{Session Overview} \label{sec:overview}
The splinter session ``Open Clusters as Laboratories for Cool Star Evolution'' was held during  
the $23^{\mathrm{rd}}$ edition of the Cambridge Workshops on Cool Stars, Stellar Systems and the Sun in Tokyo, Japan. The session brought together review material, ten contributed talks, and ten poster presentations focused on three themes of stellar rotation, magnetic activity, and chemical abundances.


\subsection{Review of Photometric and Spectroscopic Observations of Open Clusters}


\textbf{Summary:} Photometric and spectroscopic surveys of open clusters have been essential empirical anchors for studies of stellar evolution, especially regarding a star's angular momentum and magnetic activity. Over the past decades, the community has assembled rich observational databases for studying these phenomena; as the talks in this session demonstrate, that work has continued with important contributions since Cool Stars 22. Studies to characterize these phenomena fundamentally rely on catalogs of open cluster members, which are primed for revisions based on updated astrometry in Gaia's upcoming DR4 data release. \emph{Gaia}'s astrometry is sufficiently accurate and precise that the community is increasingly able to identify currently unbound former cluster members who nonetheless appear to share a common formation history. Careful attention to the ways in which explicit and implicit selection effects shape the contents of a `cluster catalog' is increasingly important for interpreting the complementary insights provided by studies of populations that remain physically bound in the present day, as well as identifying unbound structures that share a common formation history. 

\textbf{Angular Momentum Evolution:} Advances in the efficiency of measuring stellar rotation rates in benchmark open clusters, often as a byproduct of a separate scientific goal, have driven our growing understanding of stellar angular momentum evolution. In the pre-CCD era, \citet{Skumanich1972} revealed the clear signature of stellar spin-down by comparing rotation rates inferred from $v$sin$i$ measurements of Pleiades and Hyades members to that of our Sun. The development of cameras with large-format digital detectors in the mid-90s enabled new measurements of stellar rotation periods in focused, ground-based monitoring campaigns of open clusters (often motivated primarily to search for eclipsing binaries). The wealth of data generated by these programs was well illustrated by \citet{Irwin2009}, as well as the need to move beyond a pure $t^{-1/2}$ Skumanich-like spin-down toward models with mass-dependent timescales for distinct phases of angular momentum evolution: disk-locking, pre-main sequence contraction, and spin-down due to magnetized stellar winds. 

In recent years, the calibration dataset has continued to grow, adding additional benchmarks at intermediate ages and expanding coverage of the oldest stars and the critical period where stars approach the zero-age main sequence. These additions have been enabled by continued improvements in the efficiency and bandwidth of ground-based photometric monitoring surveys, such as HATnet, PTF, ZTF and MEarth, and the remarkable precision achieved by space-based exoplanet missions like CoRoT, \emph{Kepler} and TESS. The more recent compilations of rotation period measurements by \citet{Curtis2020} and \citet{VanLane2025} -- for more on the latter, see section \ref{ssec_phil} -- illustrate well the dramatic improvements in the mass and age coverage of period measurements relative to the state of the field captured by \citet{Irwin2009}. 

Notable among the discoveries enabled by this rich database of surface rotation periods is the detection of epochs of stalled spin-down for Sun-like stars \citep{vansaders2016, Metcalfe2016} and, at later dates, lower-mass K dwarfs \citep[][, and see Section \ref{ssec_chiti}]{Curtis2019a}. Though these epochs of $\sim$constant surface rotation rate were first detected observationally, as \citet{Spada2026} demonstrate, they can be explained well by either setting the mass dependence of the wind-breaking law to be proportional to the depth of the star's convective layer, or by adopting a broken power-law to govern the transfer of angular momentum from the star's core to its convective envelope. Improvements to physical models of angular momentum evolution are also being motivated at younger ages, where the short transition between surface spin-up due to pre-main-sequence contraction and the onset of spin-down due to torques from stellar winds requires dense sampling in both mass and age. Recent measurements by \citet{Douglas2024} of rotation periods in several ZAMS-age clusters illustrate this transition and constrain the mass dependence of the torque laws that begin dominating the stars' angular momentum evolution in this epoch. Specifically, \citet{Douglas2024} find that classical torque laws are not sufficient to reproduce the observed period distributions, but instead favor torque-laws similar to that advanced by \citet{Breimann2021}. 



\textbf{Stellar Magnetic Activity:} While photometric surveys are currently driving advances in our understanding of stellar rotation, systematic spectroscopic surveys of open clusters have been essential for documenting the dependence of magnetic activity on a star's mass, age, and rotation rate. While \citet{Skumanich1972} again laid the foundation for a clear physical connection between stellar rotation and chromospheric activity, richer datasets of optical spectroscopy or X-ray photometry \citep[e.g., ][]{Pizzolato2003, Pace2004} were required to reveal the distinct phases of 'saturated' and 'unsaturated' activity for rapid and slowly rotating stars, respectively.  These studies often rely on sparse sampling of cluster populations, but highly multiplexed spectrographs increasingly enable complete spectroscopic surveys of single-cluster populations. When combined with a similarly complete database of stellar rotation periods, these rich datasets offer valuable leverage for untangling the mass dependence of the age-activity-rotation relation. As an example, \citet{Frasca2025} recently provided a comprehensive spectroscopic survey of chromospheric activity throughout a single cluster, the Pleiades. Their study reveals clear differences in the slope of the relationship between a star's Rossby number and chromospheric activity level as a function of stellar mass: while hints of this mass dependence had been detected in prior studies, the dense mapping of the cluster's membership and clear removal of age as a confounding variable significantly enhance the ability to detect this signal. This rich dataset of chromospheric activity measurements in the Pleiades also enables comparisons against complementary tracers of magnetic activity, such as spot-coverage indicators revealed by the color anomalies of active K and M dwarfs \citep[see, e.g., ][ and further discussion in section \ref{ssec_cao}]{Stauffer2003, Covey2016, Cao2025} and direct measurements of surface magnetic field strengths \citep[see e.g., ][and discussion in Section \ref{ssec_wanderley}]{wanderley2024a}.  

Fully mapping the morphology of the age-rotation-activity relation across all stellar masses, however, requires connecting such surveys across clusters which capture distinct ages and metallicities. These analysis benefit from a large database of benchmark clusters, which require comprehensive spectroscopic surveys such as that recently completed by \citet{Agueros2025} for Coma Berenices, including a rich set of rotation period measurements from both ground- and space-based observatories. This survey establishes Coma Ber as a new anchor point for the age-activity-rotation relation alongside the Hyades and Praesepe, valuably expanding coverage in metallicity at this intermediate age. Homogeneous analysis of multiple clusters with the same instrument can also help alleviate systematic errors due to heterogeneous data, as \citet{Long2025} demonstrate in measuring chromospheric $R_{\mathrm{HK}}$ indices for stars in multiple clusters with spectra in the LAMOST database, inferring the slope of the age-activity relation as a function of mass.  Given the paucity of observationally accessible clusters as the oldest ages, the complementary field-star approach, such as that led by E. Pass and collaborators, discussed in section \ref{ssec_pass}, will also remain critical for constraining the age-activity-rotation relation for ages comparable to the Sun and older. 

\textbf{Updating censuses of cluster members:} In the near term, spectroscopy continues to play an indispensable role in confirming and cleaning membership catalogs based on photometric and astrometric measurements. Large homogeneous surveys such as APOGEE \citep{Guerco2025} and the \emph{Gaia} DR3 spectroscopic sample \citep{Pancino2026} add radial velocities, chemistry, and line-broadening measurements that discriminate genuine members from kinematic interlopers, refine the census of cluster membership, and provide a solid foundation of membership status for which all downstream analysis such as the activity and rotation analysis depend.

Looking ahead, however, the wealth of empirical data becoming available to searches for coherent structures in spatio-kinematic and rotation/activity spaces are making it more pressing to revisit, or at least define more explicitly, the distinction between two conceptually different notions of a "population": a system that is gravitationally bound in the present day, or alternatively a coeval population which exhibits a connected spatio-kinematic structure. 
\emph{Gaia}-era censuses are increasingly able to separate these categories quantitatively: \citet{Hunt2024} use Jacobi radii and completeness-corrected cluster masses to show that only a fraction of catalogued clusters are consistent with being gravitationally bound, and \citet{Hunt2026} characterize the selection function that governs which such systems are detectable in the first place. Against this bound-system framework sits a growing inventory of unbound, single-age structures that share spatio-kinematic properties without a surviving bound core, such as the Theia groups of \citet{Kounkel2019} and the extended complexes surrounding young clusters such as $\alpha$ Per \citep[see e.g., ][ and discussion in section \ref{ssec TARS}]{Boyle2023}.

The distinction has been sharpened by the now-routine detection of tidal tails emanating from benchmark clusters, demonstrating that currently bound clusters can nevertheless be in the process of converting into extended, unbound, coeval populations. The Hyades tails, discovered independently by \citet{Roser2019a} and \citet{Meingast2019a} and subsequently modeled dynamically \citep{Oh2020}, provide the clearest demonstration of a bound cluster shedding a substantial coeval population into the surrounding field. As such detections proliferate, so does the need to assess their reliability: \citet{Jadhav2025} examine the demographics and robustness of a large sample of candidate tidal-tail populations, flagging 40+ cluster-tail systems as robust detections and providing a framework for distinguishing a genuine detection of a dissolving cluster from kinematic contamination by interloping field stars.

Complicating matters, however, are recent detections of several dispersed systems that blur the boundary between disrupted clusters and primordial streams. Meingast 1 (also known as the Pisces–Eridanus stream) was discovered as a $\sim$400 pc coeval stream \citep{Meingast2019b}; its age was revised from $\sim$1 Gyr to $\sim$120 Myr by gyrochronology \citep{Curtis2019b} and corroborated through chromospheric activity \citep{Arancibia-Silva2020}, with membership expanded and refined by \citet{Roser2020} and \citet{Ratzenbock2020}. Theia 456 (also independently catalogued as COIN-Gaia-13) traces a similar arc: first identified as a coherent group \citep{Kounkel2019, Cantat-Gaudin2019}, it was then confirmed as physically coeval through chemistry and gyrochronology \citep{Andrews2022}, and finally shown by kinematic back-integration to have been born in a more compact state that was subsequently disrupted \citep{Tregoning2024}. OCSN 49 \citep{Hunt2023, Qin2023} has similarly been characterized as a catastrophically disrupted cluster whose core has been largely destroyed, leaving primarily a stream without a substantial present-day cluster population \citep{Miller2025} — an end-state that anchors an extreme end of the cluster disruption process.

At the limit of this continuum, and most in need of new terminology and efforts to reconcile with our existing paradigms for stellar populations, lie fully unbound structures with no identifiable cluster core at all. \citet{Ratzenbock2025} place these disk streams into the broader "fabric" of the Galactic disk, finding multiple streams with stellar densities that exceed N-body predictions by one to two orders of magnitude. The number of these streams detected in a relatively small footprint on the sky implies that coeval but unbound populations may be a generic, long-lived product of clustered star formation. The proliferation of these systems is precisely what motivates a more careful vocabulary: whether a population is defined by its gravitationally bound state or by its shared age and continuous spatio-kinematic distribution has real consequences for how it is counted, what it is calibrated against, and if/how it may be used as an age benchmark.  

To vastly oversimplify, bound clusters are likely most useful for studies of stellar evolution that require pristine and precise single-age populations, or studies of cluster populations that include detailed modelling of selection functions and/or biases imposed by the loss of smaller, less bound clusters from the detected sample. Conversely, unbound co-eval populations are likely more useful for studies of the star formation process, but such studies will require careful modelling to solidify the physical association of the coeval stars, and to interpret the population's present-day spatio-kinematic structure as informative of the star formation process and history of the Milky Way. Nonetheless, both types of definitions do hold value and promise for understanding stellar evolution and the star formation process and history in the Milky Way, even if they are not interchangeable definitions and do require more careful and explicit definitions. 











\section{Rotational Evolution} \label{sec:rotation}
Understanding the evolution of stellar angular momentum is fundamental to unveiling the magnetic histories of cool stars and establishing robust empirical age-dating techniques. Gyrochronology relies on the predictable long-term spin-down of stars driven by magnetically channeled stellar winds, where the efficiency of torque loss is intimately connected to internal structure and convective properties. However, building a predictive rotational model requires mapping the complex interplay between distinct magnetic and structural regimes from the early pre-main sequence to advanced main-sequence phases. This involves characterizing the pre-main-sequence spin-up driven by gravitational contraction, the onset of magnetic braking saturation in fast rotators, and the mass-dependent phenomena of rotational stalling. Central to these dynamics, particularly for stars between the fully convective boundary ($\sim0.35M_\odot$) and the Kraft break ($\sim1.3M_\odot$), is the degree of core–envelope decoupling. This controls how angular momentum is transported within the star and how quickly the radiative core and convective envelope exchange angular momentum. Nevertheless, stellar rotation alone provides limited constraints on stellar ages, particularly below $\sim 80$~Myr \citep{Bouma23_emp_lim}, where the broad dispersion of rotation periods among coeval stars limits the precision of gyrochronology. Furthermore, the mass-dependent evolution of rotation is not monotonic: low-mass stars can undergo extended periods of stalled spin-down after converging onto the slow-rotator sequence, as revealed by observations of Praesepe, NGC 6811, and NGC 752 \citep{Curtis2019b, Curtis2020}.
 
The following subsections summarize the latest observational constraints and theoretical models presented during this session.

 
\subsection{Age inference from photometric light curves}\label{ssec_phil}

\begin{figure}[ht!]
    \centering
    \includegraphics[width=0.95\linewidth]{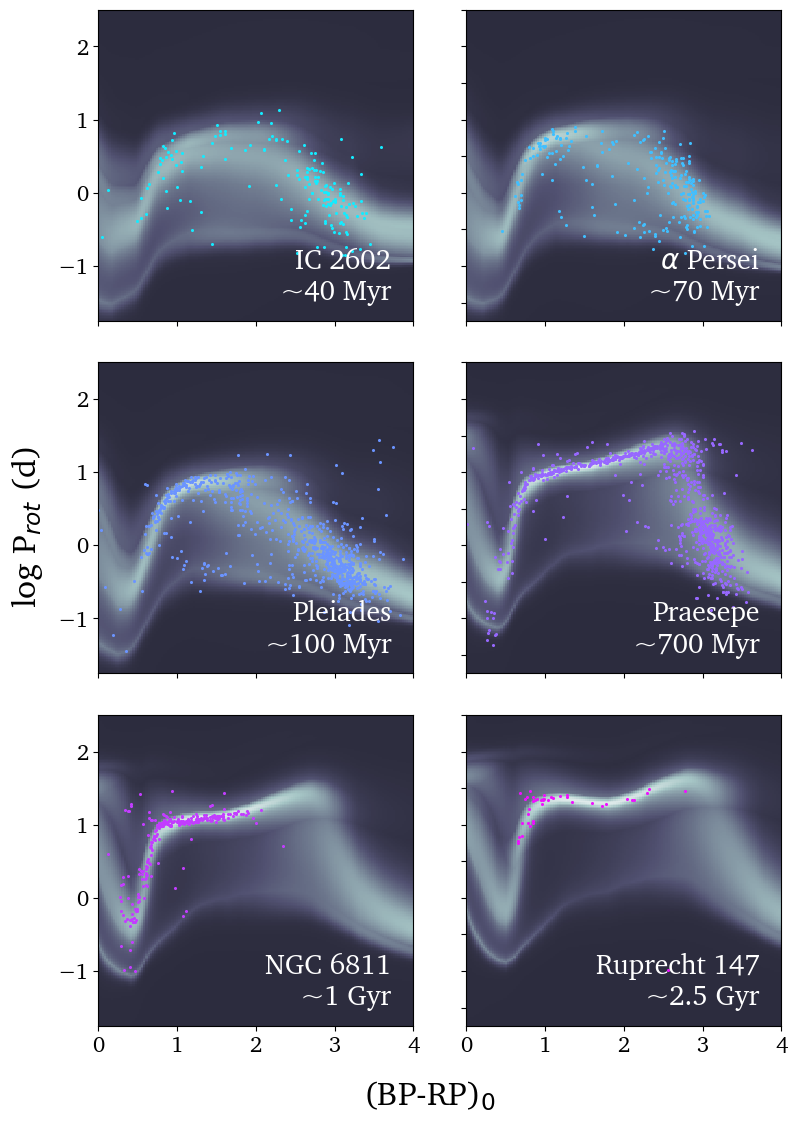}
    \caption{Six of the open clusters used to develop \texttt{ChronoFlow} \citep{VanLane2025}. Background shading represents the expected density of stars in $(BP-RP)_0 - P_{rot}$ space from \texttt{ChronoFlow} at the cluster ages (lighter = higher density), and colored points are cluster members.}
    \label{fwfigure1}
\end{figure}

As described previously, open clusters are extremely useful for calibrating gyrochronology models since they provide independent stellar age estimates. Here we present \texttt{ChronoFlow} \citep{VanLane2025}: a fully continuous and probabilistic data-driven gyrochronology model, calibrated on rotators from 30 open clusters and associations that were compiled from a deep literature review. \texttt{ChronoFlow} implements a \textit{Conditional Normalizing Flow} architecture, which enables complete flexibility in modeling rotation period distributions (see Fig. \ref{fwfigure1}). Importantly, it also mitigates the selection effects in color and age, which can be significant in open cluster observations.

As part of this work, we characterized various systematics important to gyrochronology, such as choice of de-reddening procedure, cluster membership cuts, and uncertainty in calibration ages. \texttt{ChronoFlow} can be used to forward model the rotational evolution of stars, and infer ages of individual stars or populations from $(BP-RP)_{\mathrm{0}}$ and $P_{\mathrm{rot}}$ measurements. It is publicly available at \href{https://github.com/philvanlane/chronoflow}{https://github.com/philvanlane/chronoflow}.

In this talk, we also discussed initial results from a \textit{Time Series Foundation Model} (TSFM) that we have developed: \texttt{EncoTESS}, which is now publicly available at \href{https://github.com/philvanlane/encotess}{https://github.com/philvanlane/encotess}, and the associated paper has been submitted \citep{VanLane2026_EncoTESS}.

Conceptually, TSFMs are intended to use raw time series data directly as input to learn important characteristics of the underlying processes, instead of relying on summary statistics. With \texttt{EncoTESS}, our goal is to maximize the information we can extract from \textit{TESS} light curves to improve age constraints over gyrochronology alone.

We show that \texttt{EncoTESS} can recover open cluster ages more accurately and precisely for young red stars that have not converged onto the slow rotator sequence yet, compared to using rotation period or variability amplitude as indicators. This demonstrates the ability of TSFMs to extract age information from light curves that is \textit{complementary} to rotation.

\subsection{Spindown of fully convective M dwarfs} \label{ssec_pass}

Fully convective M dwarfs spin down differently from the gradual, Skumanich-law spindown of Sun-like stars, as evidenced by the bimodal rotation periods observed for these dwarfs in the field \citep{Newton2016, Newton2018}. Some rotate quickly, with periods shorter than 2 days. Others rotate slowly, with periods longer than 100 days. There are few with intermediate periods, implying that the transition between these modes is abrupt. However, existing gyrochronological relations \citep[e.g.,][]{Engle2023} do not reproduce this behavior, instead predicting few short-period rotators, many intermediate rotators, and many slow rotators. So, how do fully convective M dwarfs actually spin down? While age estimates are challenging for M dwarfs, we can use open clusters, wide binaries, and a volume-complete survey of our nearest neighbors to uncover their rotational evolution.

The oldest clusters in which rotation periods have been measured for such small stars are Praesepe and the Hyades ($\sim$700~Myr). We show that at this age, nearly all fully convective M dwarfs are still rapidly rotating; this result was not obvious in previous work \citep{Douglas2019} due to systematic biases in stellar mass estimates \citep{Pass2022}. Using our other samples, we find that fully convective M dwarfs spin down gradually (P$\sim$0.1--10 days) for 1--4~Gyr (depending on mass), then abruptly transition to long periods (P$>$70 days) over a timescale of $<$100~Myr \citep{Pass2022, Pass2023, Pass2024}. The dispersion in the mass-dependent spindown relation is small ($<$0.5~Gyr) when stellar birth environment, age, and metallicity are controlled, but we have identified some M dwarfs that make the transition at much younger ages—systems that may represent special initial conditions (e.g., high-energy birth environment and initial rotation rate) that led to a shorter saturation lifetime \citep{Pass2022, Pass2024}. In addition to offering a unique parameter space for testing theoretical spindown models, our findings have important implications for planetary atmosphere retention and the ‘cosmic shoreline’ of fully convective M dwarfs, suggesting that many key planets for atmospheric characterization received a historic XUV flux that exceeds the canonical scaling relation by more than a factor of 3 \citep{Pass2025}.
 
\subsection{The Greater Pleiades Complex and TESS rotation surveys}
\label{ssec TARS}

\begin{figure}[ht!]
    \centering
    \includegraphics[width=0.95\linewidth]{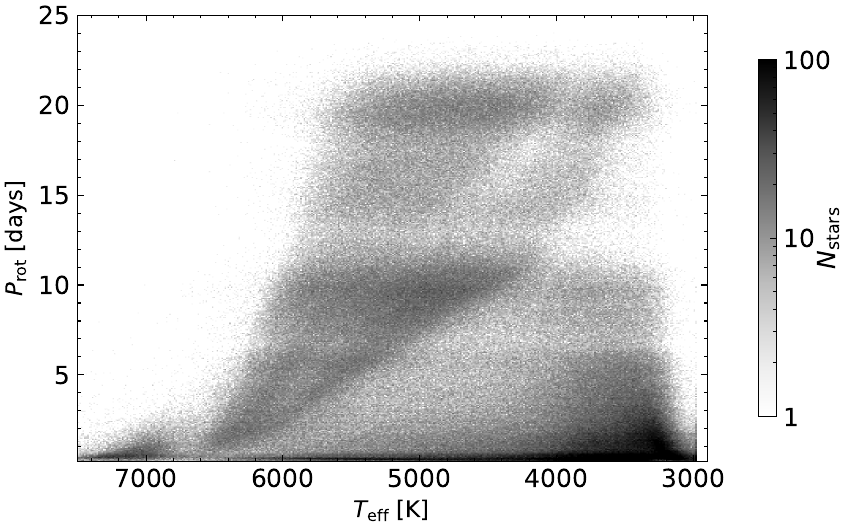}
    \caption{The rotation period vs. effective temperature distribution of the TARS sample. TARS contains periods up to 25 days for more than one million stars, and we estimate that 93\% of these periods are rotation periods.}
    \label{fig:tars}
\end{figure}

Stellar rotation provides an empirical probe of stellar age, magnetic activity, and angular-momentum evolution. The all-sky coverage of TESS makes it possible to extend rotation measurements beyond individual clusters to perform a uniform search for stellar rotation in bright stars near the Sun. We developed the \textit{TESS All-Sky Rotation Survey} (TARS) to exploit this capability using the TESS full-frame images. TARS provides a flux- and distance-limited survey of $\sim8$~million stars within 500\,pc with $T<16$, with light curves corrected for instrumental systematics using a Causal Pixel Model and candidate rotation periods measured using Lomb--Scargle periodograms. A central goal of the survey was to quantify not only the measured periods, but also their reliability and the survey completeness. The default TARS catalog provides period measurements for more than one million stars (see Fig. \ref{fig:tars}), along with multiple catalog layers with increasingly restrictive quality criteria, enabling users to construct their own catalogs that trade completeness against reliability according to the needs of a given application. We additionally presented a method to identify and correct the half-period harmonics that are inherent to rotation period measurements from TESS data, thereby allowing us to recover rotation periods as long as 25 days from a single TESS sector. The resulting catalog provides a homogeneous view of stellar rotation across the local Galaxy and enables rotation to be used for identifying young stellar populations and both stellar and exoplanetary astrophysics.

One immediate application of TARS is the search for the dispersed remnants of young clusters. Spatio-kinematic clustering searches with \emph{Gaia} are highly effective for identifying compact structures in position--velocity space, but become increasingly incomplete as associations disperse into the Galactic field. Rapid rotation provides an independent indicator of youth that can recover contrast in this regime. Combining the TARS rotation measurements with \emph{Gaia} kinematics, we developed a Bayesian framework that evaluates stellar membership using both rotation and phase-space information. Applying this method to the region surrounding the Pleiades revealed the \textit{Greater Pleiades Complex}, a spatially extended and kinematically connected population consistent with having originated from the same broader star-formation event as the Pleiades. The complex encompasses multiple previously identified groups, including the AB Doradus moving group, demonstrating that rotation-selected samples can reveal coherent stellar populations even after their spatial and kinematic signatures have become diffuse. This application illustrates the broader utility of TARS: beyond providing rotation periods for individual stars, the survey enables systematic searches for the dispersed structure of recent star formation throughout the solar neighborhood.

\subsection{Nearby new Open Clusters with Gaia}

Open clusters and coeval stellar associations provide essential benchmarks for studies of stellar rotational evolution and gyrochronology. However, the number of nearby populations with well-characterized rotation sequences remains limited, leaving parts of the age--rotation relation sparsely sampled. Using hierarchical clustering of \textit{Gaia} EDR3 and DR3 astrometry, we identified hundreds of previously unreported nearby stellar clusters, associations, and extended structures within approximately 200 pc.

To characterize their rotational properties, we developed an automated Gaussian-process (GP) framework to infer stellar rotation periods from TESS photometric time series. GPs are well suited to this problem because rotational modulation is generally quasi-periodic, reflecting the evolution of starspots and active regions. Their flexibility, however, can also lead to overly complex solutions that reproduce short-timescale light-curve structure without recovering the dominant rotational signal.

We therefore investigated an entropy-inspired regularization of the GP likelihood designed to reduce unnecessary model complexity while preserving the ability to describe evolving stellar variability. The method was validated on several hundred stars in well-characterized young populations with literature rotation periods. Moderate regularization improves the recovery of established periods, reduces harmonic and overly complex solutions, and increases the consistency of inferred periods between independent TESS sectors. Excessive regularization can suppress genuine stellar variability, indicating that the best performance occurs in an intermediate regime between model flexibility and interpretability.

This framework provides a homogeneous approach for measuring rotation periods in the newly identified \textit{Gaia} populations. Their color--period sequences can then be compared with established stellar benchmarks to constrain their relative ages and placement within the empirical sequence of rotational evolution. Together, these new populations and the regularized GP methodology expand the set of nearby benchmarks available for studies of stellar rotation, age, and magnetic evolution. The study describing the regularized GP methodology has been submitted and accepted subject to revision (Moranta et al. 2026, \textit{submitted}), while the characterization of the newly identified stellar populations with \textit{Gaia} DR3 data is currently in preparation (Moranta et al. 2026, \textit{in prep}).

\subsection{Do K dwarfs really stall? Zeeman-Doppler imaging of the Hyades} \label{ssec_chiti}

Cool stars lose angular momentum over their lifetimes through magnetized winds, producing a tight relationship between rotation period and age. At intermediate ages ($0.6$--$3$\,Gyr), however, K-dwarfs in open clusters show an apparent stalling in this spin-down, challenging the models on which gyrochronology depends and exposing gaps in our understanding of angular momentum evolution in cool stars. 

Two hypotheses have been proposed: the braking torque is significantly reduced in these stars, or internal angular momentum transport from the core replenishes the surface loss (core-envelope recoupling). These hypotheses predict opposite behavior for the large-scale dipole field and braking torque, which we test directly using the first Zeeman Doppler Imaging (ZDI) maps of K-dwarfs in the stalling regime, reconstructed from LBT/PEPSI spectropolarimetry of Hyades stars. 

We find that dipole fields, the main drivers of magnetic braking, dominate the field morphology in all stars, including stalled K-dwarfs, with strengths that scale with Rossby number as expected. Stalled stars also show elevated starspot covering fractions, mirroring the enhanced activity previously reported in stalled Praesepe K-dwarfs. Combining ZDI-derived field strengths with X-ray-based mass-loss rates, we compute instantaneous braking torques that agree with standard model predictions across the full sample, including the stalling regime. Reconciling a standard, solid-body model with the stalling observed between Hyades/Praesepe and NGC~6811 age instead requires suppressing the torque by two to three orders of magnitude, unsupported by the observed torques even after accounting for mass-loss rate uncertainties and known ZDI biases. 

Therefore, our results disfavor a suppressed braking torque as the cause of spin-down stalling, favoring internal angular momentum transport instead.

\subsection{Stellar rotation across seven young open clusters within the PLATO LOPS2 field}
Early star-disk interactions act to prevent stars from spinning up as they contract along the pre-main sequence, in a process known as disk locking \citep{Koenigl1991, Rebull2018}. A distribution of disk lifetimes for a given mass results in a spread of time-frames for which a star can spin up, which is further propagated into a spread of rotation periods for any given mass upon arrival at the zero-age main sequence. Once on the main sequence, contraction is halted, and stars lose angular momentum via stellar winds, eventually converging onto a well-defined mass-dependent rotation sequence that evolves monotonically as a function of age. Observations of open clusters with well-defined ages, therefore, provide the ideal laboratories to study these mechanics, allowing calibration of the empirical gyrochronology models that track these evolving sequences \citep[e.g.][]{Bouma23_emp_lim, VanLane2025}. \par

In our recently published work \citep{Hughes2026}, we present rotation in seven young open clusters, spanning ages $\sim$ 40--700\,Myr, that reside in the upcoming PLATO LOPS2 field. For three of the clusters: Alessi 3, Trumpler 10 and NGC 2451\,B, the rotation distributions mark the first comprehensive period distributions to date, whilst we also expand existing distributions for NGC 2451\,A, NGC 2516, Collinder 135 and IC 2391. In total, we report 1063 rotation periods, of which 479 are newly analysed as part of cluster-specific rotation studies, and 63 are not included in the recently published TESS All Sky Rotation Survey (TARS: \citep{tars_Aboyle}, refer to Section \ref{ssec TARS}). Of the 1063 reported rotation periods, we identify 285 as likely higher-order star systems and thus remove them from consideration when analyzing rotational evolution of single star systems.\par 
We find that for main-sequence solar-mass stars in the $\sim$40\,Myr NGC 2451\,B, the slow sequence has already converged and is distinguishable from the current canonically accepted lower gyrochronal age anchor of $\sim$80\,Myr set by $\alpha$ Persei \citep{Boyle2023}. Thus, these results significantly lower the age at which open clusters can be relatively aged via rotation. Furthermore, we identify stalled spin down for stars $\gtrsim$\,$1$\,$M_{\odot}$ in NGC 2451\,A to at least the age of NGC 2516 ($\sim$70--150\,Myr), supporting recent suggestions that stalled or slowed spin down should be observed at a variety of ages and masses \citep{Spada2020}, and suggest that slowed or stalled spin down propagates through rotation-mass space as a function of both mass and time. Finally, we utilise differential gyrochronology aging to provide an age estimate for Alessi 3 of 687$\,\pm$\,106 Myr, significantly lowering the spread in the current literature age estimates for the cluster that have previously been restricted to isochronal fitting.

\subsection{Pushing Gyrochronology to younger ages using open clusters}

We present TESS rotation periods for 7 young open clusters ($\gamma$~Vel, 25~Ori, NGC~2232, IC~2391, IC~2602, IC~4665, ABDMG). Combined with 7 literature clusters (NGC~2264, NGC~2547, NGC~2541A, $\alpha$~Per, Pleiades, Blanco~1, NGC~2516), our sample spans ages from $\sim$\,3--150 Myr and reveals a coherent mass-dependent transition from stellar spin-up to spin-down. 
To trace the evolution of stellar rotation in FGK stars, we identify two rotational populations within $T_{\rm eff}$ bins: a slow-rotator population defined by the 80th-percentile rotation period, and a fast-rotator population forming a more scattered branch. These two components allow us to trace the competing effects of stellar contraction and magnetic braking as a function of effective temperature. In particular, this approach accounts for the fact that stars of different masses contract and evolve on different timescales, reaching the Main Sequence at different ages (Caccherano et al. \textit{in prep}).
Extending gyrochronology to ages younger than the current $\sim$80~Myr anchor \citep{Bouma23_emp_lim} is crucial for establishing stellar angular-momentum evolution as a more powerful age and magnetic-activity diagnostic for young stars in open clusters. However, additional age-sensitive diagnostics, such as photometric variability amplitude, can provide independent age constraints and help break the rotation--age degeneracy at ages where stars are still converging towards their slow rotator sequences.


\section{Stellar Magnetic Activity} \label{sec:activity}

Stellar activity and rotation are closely linked phenomena in cool stars coupled by the stellar dynamo. Activity signatures from the photosphere (e.g. photometric spot modulation amplitude), through the chromosphere (e.g. emission in spectral lines such as H$\alpha$ or Ca II H\&K), to the corona (e.g. soft X-ray emission) trace the response of the stellar atmosphere to the magnetic field. Understanding the evolution of these signatures gives direct insight into the stellar dynamo and its short- and long-term evolution. Canonically, the two regimes of stellar activity are the saturated (associated with fast rotators) and unsaturated regimes (slow rotators). Similarly to rotational evolution (see Section \ref{sec:rotation}), detailed observations of open clusters and field stars alike show that this simple picture needs to be revised. The known rotational transitions from fast-to-slow rotators, along with periods of spin-down stallation, leave imprints in the stellar activity \citep{Fritzewski2021b, Frasca2025, Yang2025, Gody-Rivera2026}.

The following subsections present some of the latest advances in the field and their applications to open clusters.
 
\subsection{Magnetic fields of M dwarfs across cluster ages} \label{ssec_wanderley}

M dwarf stars comprise around 70$\%$ of the stars in the Milky Way, making them very important for the understanding of our galaxy. They also have higher probabilities of hosting rocky exoplanets and are more sensitive to exoplanet detection techniques such as radial-velocity and planetary transits. These properties make M dwarfs especially relevant in the search for potentially habitable worlds. However, their lower masses result in longer spin-down timescales, allowing M dwarfs to sustain strong magnetic activity over longer periods. Their magnetic fields can drive stellar winds and produce non-thermal emission, potentially affecting the atmospheres of orbiting exoplanets and contributing to atmospheric erosion. Therefore, modeling the magnetic fields of M dwarf stars is crucial not only for understanding the physics and evolution of these stars, but also for assessing their impact on the atmospheres and habitability of their exoplanets.

We used APOGEE near-infrared high resolution (R$\sim$22,500) spectra to derive mean magnetic fields of partially convective M dwarf stars (3400$<$T$_{\rm eff}$$<$4000 K) from the Pleiades open cluster (62 stars presented in \citet{wanderley2024a}), the Hyades open cluster (18 stars to be presented in Wanderley et al. 2026, in preparation), and M dwarfs from the galactic field known to host exoplanets (29 M dwarfs presented in \citet{wanderley2024b}), totalizing 109 stars. We developed a pipeline that employs Monte Carlo and Markov Chain to derive magnetic filling factors along with stellar metallicities by modeling the Zeeman broadening in four Fe I lines that are sensitive to magnetic fields (i.e., have high Landé-g factors). The pipeline uses the Synmast radiative transfer code \citep{kochukhov2010}, which considers the effects of magnetic fields in the stellar spectra, along with MARCS atmosphere models \citep{gustafsson2008}. 

\begin{figure}[ht!]
    \centering
    \includegraphics[width=0.95\linewidth]{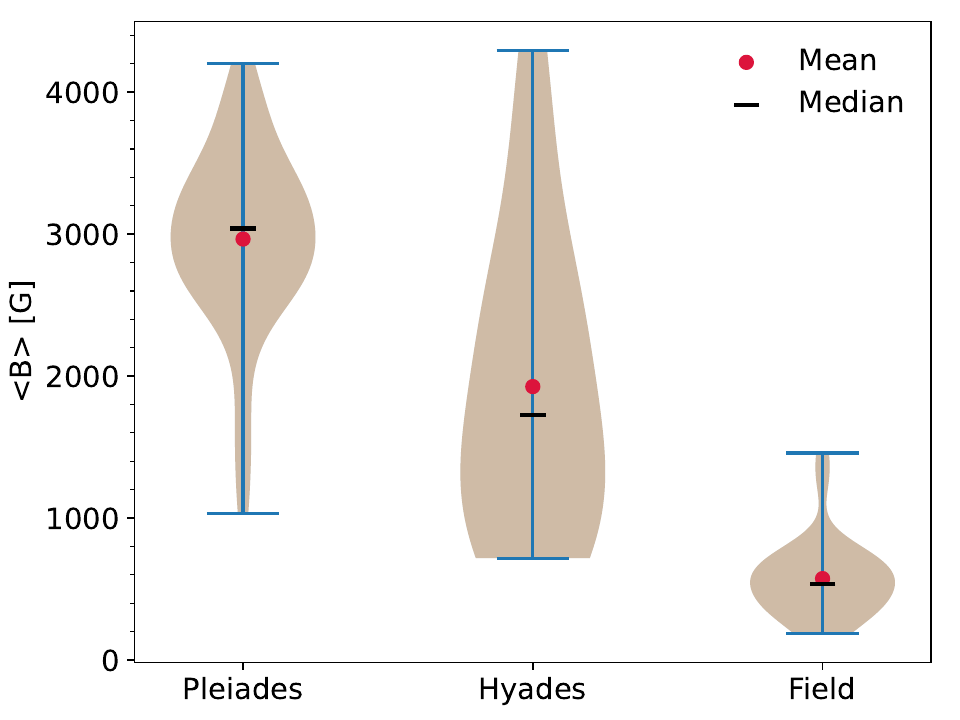}
    \caption{Violin plots of the derived magnetic field strengths for the Pleiades, Hyades, and field-star samples. The red circles indicate the mean magnetic field strength of each sample, while the black horizontal lines indicate the corresponding median values.}
    \label{fwfigure2}
\end{figure}

The mean magnetic field strengths found for the sample range from $\sim$0.2 kG to $\sim$4.3 kG. For the same mass, younger stars are generally expected to exhibit stronger magnetic fields, as they lose angular momentum and spin down over time. The mean magnetic fields measured for the stars in the Pleiades open cluster are, on average, significantly stronger (mean $\pm$ STD = $3.0\pm0.6$ kG) than those of the Hyades cluster (mean $\pm$ STD = $1.9\pm1.0$ kG). This is a result of the age difference between the two clusters, with the Pleiades stars being roughly six times younger than the Hyades stars (Pleiades and Hyades clusters have ages of $\sim$0.1 Gyr and $\sim$0.6 Gyr, respectively). The distribution of magnetic field strengths as a function of rotational period and Rossby number further supports this interpretation. While most Pleiades stars are in the saturated regime, the Hyades stars occupy both the saturated and unsaturated regimes, presenting a larger magnetic field scatter. Although we don't know the ages of the planetary host stars, since they are from the galactic field, we expected them to be some Gyr old, much older than the other two samples. This age difference is reflected in the much smaller fields of this sample (mean$\pm$STD=0.6$\pm$0.3 kG), with all stars being located in the unsaturated regime, having lost much of their original angular momentum and magnetic fields. We also investigated the distribution of mean magnetic field strength as a function of $T_{\rm eff}$ for each sample. The Hyades stars exhibit a clear trend in which cooler stars tend to have stronger magnetic fields, whereas this trend is much less pronounced in the Pleiades sample, because most of its stars are in the saturated regime. The field-star sample, in contrast, exhibits a large scatter in magnetic field strengths, which may result from the broad range of stellar ages represented in this subsample. Figure \ref{fwfigure2} shows violin plots for the magnetic fields of the three samples, together with their respective mean and median values. The figure illustrates the decrease in average magnetic field strength with increasing stellar age, as well as the larger scatter in magnetic field strengths observed for the Hyades stars.

Magnetic fields can inhibit convection in the stellar interior and generate large starspots on the photosphere. Both effects can reduce the effective photospheric temperature and cause the star to inflate. We quantified the radius inflation of the stars in our Pleiades and Hyades samples by comparing our derived radii with predictions from the MIST \citep{choi2016}, Dartmouth \citep{dotter2008}, and SPOTS \citep{Somers2020} isochrones. We also used the SPOTS isochrones to estimate the photospheric spot coverage fraction. We find a clear relationship between magnetic activity, radius inflation, and spot coverage, with more active stars exhibiting greater levels of inflation and larger spot fractions. Magnetic activity can also heat the stellar chromosphere and corona, producing enhanced non-thermal emission. To investigate the chromospheric and coronal emission of our targets, we cross-matched our Pleiades and Hyades samples with literature measurements of H$\alpha$ equivalent widths and X-ray-to-bolometric luminosity ratios, respectively. We find a clear correlation between mean magnetic field strength and non-thermal emission, with stars hosting stronger magnetic fields exhibiting, on average, stronger H$\alpha$ emission and higher X-ray-to-bolometric luminosity ratios. We also identify distinct saturated and unsaturated regimes, following the same general behavior observed in the distribution of magnetic field strength as a function of rotational period.

Our results reveal clear relationships between mean magnetic field strength and stellar age, non-thermal emission, radius inflation, and photospheric spot coverage. To date, only a limited number of M dwarf stars have precise measurements of their mean magnetic fields. Increasing the number of such measurements is therefore essential for improving our understanding of the magnetic properties and evolution of these stars, as well as for investigating in greater detail how their magnetic activity shapes the space environment of orbiting exoplanets. The full set of results presented in this work will be published in Wanderley et al. (2026, \textit{in prep}).

\subsection{Starspot evolution in open clusters} \label{ssec_cao}

Surface magnetism and its observable manifestations on cool stars evolve dramatically on evolutionary timescales \citep{Skumanich1972}. In turn, the evolution of stellar activity biases the accurate recovery of stellar parameters, raising tensions with stellar models---active stars are preferentially inflated and appear cooler than expected from standard stellar theory \citep{Cao2025}. Accurate stellar models are critical inputs for many fields within astrophysics; however, a lack of an empirical scaling for starspot evolution has so far prevented evolutionary models from consistently accounting for the effects of activity in young, cool, or active stars. Therefore, identifying the magnetic starspot evolution law as a function of stellar mass is essential for developing better stellar evolutionary prescriptions and reducing modeling systematics in low-mass stars.

We use the starspot retrieval framework developed by \citet{Cao2025} and augment their sample with open cluster stars with APOGEE near-infrared spectra (R$\sim$22,500) ranging three decades in age, from $\lambda$ Orionis ($\sim$5 Myr) to M67 ($\sim$4 Gyr), as well as field stars from the broader LEOPARD sample from Cao et al. (2026, in preparation). We fit power laws characterizing the time-evolution of the starspot filling factor in discrete $T_{\mathrm{eff}}$ intervals for our sample of open cluster members; these then define magnetochronal age relationships, which we then use to recover empirical isochrones from our field sample.

By implementing the derived empirical scaling in the public Yale Rotating Evolution Code (YREC) \citep{Pinsonneault2026}, we modify the SPOTS models \citep{Somers2020}---defined at fixed $f_{\mathrm{spot}}$---to include a time-varying starspot evolution case. We find that the characteristic systematics attributed to stellar magnetism are maximized in specific regions of the HRD where stars are known to be active: in the pre-main sequence and for cool stars, particularly the K \& M dwarfs. As a result, accounting for starspot evolution is essential for the accurate characterization of cool stars, especially for K \& M dwarfs, which can experience a prolonged span of magnetic activity \citep{Cao2023}.

The recovered empirical starspot--age relations indicate a clear mass-dependent evolutionary decline with applicability over three dex in age, featuring a faster decline for the higher-mass stars. This approach promises a complementary spectroscopic age estimator for large stellar samples, using magnetochronal relationships. The subsequent age-tagged empirical sequences appear to match well independently with magnetic and spotted evolutionary models, suggesting that deviations from standard stellar isochrones in clusters or field contexts can be interpreted as the effects of the changing stellar dynamo over evolutionary timescales. We present our final set of these magnetochronal laws and evolutionary isochrones in Cao et al. (2026, \textit{in prep}).

\subsection{Age of the Volans-Carina Moving group}
Accurate and precise ages are necessary to benchmark all stages of stellar evolution. However, measured ages are sensitive to both the method used and the choice of evolutionary model, especially for young, magnetically active stars. 
Low-mass PMS stars are larger and cooler than expected based on standard models, leading them to appear younger than high-mass stars within the same association on a CMD \citep{feiden_magnetic_2016}. These discrepancies between standard and magnetic models have been observed in low-mass stars in associations up to 50 Myr. However, it is much less well studied in older associations, with only a few associations in the age range of $100-200$ Myr.
There are also differences between measured ages using different methods, such as isochrone CMD fitting and Lithium Depletion Boundary (LDB), even when using the same models. Ages measured using LDB are up to a factor of $2$ older than ages measured using CMD fitting of low-mass stars.
The Volans Carina association (VCA) provides an excellent opportunity to extend studies of age measurements into the older age range. VCA is a nearby ($\sim 85$ pc), newly discovered association, with a membership list expected to be complete down to spectral type M7.
In this work, we use new spectra of $28$ low-mass VCA candidate members to evaluate their membership, measure the association age using the LDB method, and evaluate the fit between the association CMD and standard and magnetic evolutionary models.

Spectra were taken using the Magellan Echelle Spectrograph on the Magellan 6.5m telescope at Los Campanas Observatory. Standard calibrations were taken, and data were reduced using the Pypeit reduction pipeline \citep{pypeit:joss_pub, pypeit:zenodo}. We measured stellar RV's using cross-correlation. We used the BANYAN $\Sigma$ Bayesian analysis code, with our measured RV's and \textit{Gaia} astrometry to determine membership probability for each of the $28$ stars in our sample. We then measure the 6708\AA\, Li line in each of our confirmed stars to determine Li abundance and measure the age of the association.

Of the $28$ observed candidate members, we confirm the membership of $23$, requiring a $>95\%$ membership probability and a secondary indicator of youth (such as H$\alpha$ emission) to consider them confirmed. None of the observed confirmed or candidate members, down to a magnitude of $K_S = 9.01$, show any Li absorption, making a true LDB age impossible to determine. However, we are able to set a lower limit on the age of the association by comparing the magnitude of the lowest mass candidate and confirmed members against stellar evolutionary models. We find that VCA is at least $50\%$ older than suggested by prior isochronal age measurements \citep{gagne_volans-carina_2018}, with a conservative lower limit of $135\pm23$ Myr. A less conservative estimate, which includes high-likelihood candidate members as well as confirmed members, provides a lower limit of $163\pm35$ Myr. We also find that none of the stellar evolutionary models tested were able to fully fit the color-magnitude sequence of the association. Both standard and magnetic models systematically underestimated the brightness of low-mass stars, while magnetic models also overestimate the brightness of mid-mass stars. Overall, the significantly older age found with LDB, and the discrepancy between the association's color-magnitude and that predicted by standard and magnetic models, suggest that the age tension problem extends into older associations.

\subsection{Photometric Activity Cycles in Open Clusters} \label{ssec_chahal}
Magnetic fields generated in stellar interiors undergo cyclic regeneration, similar to the Sun’s 11-year cycle, producing starspots, flares, and other surface phenomena. Open clusters, with well-constrained ages, provide ideal benchmarks to trace the evolution of these activity cycles. We compile a sample of single stars in young open clusters (<1 Gyr) to investigate how magnetic activity and its cycles depend on age and mass. To probe long-period activity cycles analogous to the 11-yr solar cycle, we develop a method that combines time-series photometry from \emph{Kepler} FFIs \citep{Montet2017}, ASAS-SN, and ZTF. Applying this technique, we previously detected cycles in 138 fast-rotating G–K stars, revealing young Sun-like stars in the intermediate region where our Sun lies and challenging the traditional active–inactive branch paradigm \citep{Chahal2025}. We now extend this analysis to open clusters observed in \emph{Kepler} FFIs spanning 50 Myr to 1 Gyr, detecting photometric cycles in over 100 cluster members (Chahal et al., \textit{in prep}). We find that the cycle amplitude is higher for longer cycle periods, suggesting that shorter cycles are driven by near-surface dynamos, while longer cycles likely originate from deeper convective or interface-layer dynamos. Most detected cycles occur in younger, fast-rotating stars, with a marked decline in the detection of longer cycles among older stars. We also identify several young stars lying above the traditional active branch in the cycle–rotation diagram, challenging its validity, particularly at young ages.

\section{Chemical Abundances} \label{sec:abundances}
Open clusters are typically assumed to have formed from a chemically homogeneous cloud of gas and dust. Together with their broad coevality, stars in these clusters are ideal for testing chemical clocks. With the advent of high-resolution spectroscopic surveys (e.g. GALAH, Gaia-ESO, SDSS/APOGEE), the required elemental abundances are accessible on a Galactic scale. These surveys provide the data for more traditional clocks such as lithium depletion, as well as individual abundance ratios (e.g. [C/N], [Y/Mg]) that trace the Galactic enrichment.

The following subsections present some of the latest results for open clusters based on these current spectroscopic surveys.
 
\subsection{Testing chemical clocks with subgiants} \label{ssec_myers}

Stellar abundances encode the chemical enrichment history of the Milky Way, which has progressed over successive generations of star formation. This inherent relationship between abundance and stellar age (or chemical clock) has been proposed as a hopeful, relatively cheap alternative to traditional age-dating methods, especially with the plethora of large-sky spectroscopic surveys available. However, the reliability and generalizability of these chemical clocks remain uncertain. 

In this work, we use the precise and accurate subgiant age catalog of \cite{nataf_2024} to systematically evaluate the predictive power of all possible linear combinations of GALAH DR4 \cite[][]{galah_dr4} abundances. We first split the data into high- and low-$\alpha$ populations and use cross-validation and the Bayesian Information Criterion (BIC) to subsequently evaluate each model. 

Of the more than 100 million models tested, we find a significant correlation between age and chemical abundances in the low-$\alpha$ population, but no comparable relationship in the high-$\alpha$ population. The high-$\alpha$ population is known to be older on average, and the absence of a measurable abundance–age relation may reflect the heterogeneous chemical histories of its stars. In contrast, the significant abundance–age correlations observed among low-$\alpha$ stars suggest a more coherent chemical enrichment history that can be traced in the solar neighborhood.

Our statistically preferred chemical clock includes eight abundances spanning a range of nucleosynthetic families, from light elements to heavy neutron-capture elements. This suggests that accurately predicting stellar ages requires information from a diverse set of chemical abundance families rather than relying on a single abundance ratio.
This clock achieves an RMSE of 0.8 Gyr; however, accounting for uncertainties in both the coefficients and abundances increases the average age error to 1.3 Gyr. To evaluate the applicability of this model to broader stellar populations, we apply our statistically preferred chemical clock to five open clusters with ages ranging from 2.3 to 5.8 Gyr, each with more than five stars with reliable GALAH abundances. Open clusters provide an ideal test of our relationship because their stars are coeval and chemically homogeneous. After applying our selection criteria, the remaining cluster stars are all red giants.

The predicted ages exhibit large scatter and significant offsets from the known cluster ages, indicating that the chemical clock does not generalize from subgiants to red giants in these clusters. More broadly, although chemical clocks offer a potentially efficient and inexpensive means of estimating stellar ages, our results demonstrate that abundance–age relationships may not be universally applicable across different stellar populations and evolutionary stages.
Taken together, these results suggest that the existence of a strong abundance–age correlation within a particular stellar population does not necessarily imply that the relationship can be generalized to other populations or evolutionary stages.

 
\subsection{Neutron-capture element gradients}

Open clusters are ideal objects for analyzing variations in abundance with both location and time, because they are populations that have the same chemical abundances, velocities, distances, and age. In this work, we leverage these characteristics to analyse the Galactic abundance gradients for 22 different abundances, ranging from the alpha elements to neutron capture. As a basis for our observations, we use the APOGEE DR17-based open cluster catalog: \citet{myers_2022}, supplemented with members from \citet{Cantat-Gaudin2020b}. We follow up selected stars with high-resolution, high-S/N, optical spectra from the MIKE spectrograph at Las Campanas Observatory or the HIRES instrument on the Keck telescope. 

Our final sample consists of 57 stars in 18 different open clusters. We derive the final abundances with the Brussels Automatic Code for Characterizing High-accUracy Spectra   \cite[BACCHUS;][]{masseron_2016}. For each abundance that overlaps with APOGEE DR17, we find good agreement between the gradients measured in \cite{myers_2022} and those measured in this work. The addition of the neutron capture elements also revealed a trend in the slopes of the Galactic abundance gradient as a function of s-process enhancement. That is, the abundances with significant s-process enrichment (Y, Sr, Zr, Ba, La, Ce) show significantly shallower slopes than those which do not have significant s-process enrichment.

 
 

\section{Panel Discussion} \label{sec:panel}

The panel discussion with Kevin Covey, Stephanie Douglas, David Montes and Loredana Prisinzano focused on the challenges of using stellar rotation and Li depletion as probes of stellar evolution and age, with questions from the audience focusing on the interpretation and limitations of open-cluster observations, as well as the kinds of future observations needed to make progress in calibrating stellar evolution.

The first main theme concerned the representativeness and reliability of the benchmark clusters currently underpinning empirical studies, particularly rotation studies. Although clusters such as the Pleiades, Praesepe, and NGC 6811 provide some of the best available rotational datasets, questions have been raised about whether their properties are representative of the wider stellar population. Sparse age sampling combined with differences in cluster metallicity, observational and membership completeness, and criteria adopted to clean cluster samples of field interlopers and distinguish single-star members from binaries and multiples may introduce systematic effects in the observed rotation--colour relations and their physical interpretation. These concerns also extend to the absolute ages assigned to clusters, for which systematic offsets between different age-determination methods (e.g. gyrochronology, Li depletion boundary, standard vs. starspot-influenced magnetic isochrones, etc) could propagate directly into empirical relations.  In particular, accounting for the effects of magnetic spots in stellar models is crucial to obtaining realistic and consistent isochronal ages \citep[e.g.][]{tarantino2025}. More generally, the discussion highlighted the importance of consistent approaches to membership and multiplicity when constructing benchmark samples for studies of stellar evolution and age. 

A second theme was the value of combining different types of observations for the same stars. Rotation periods, spectroscopic and chemical information, and kinematics can provide complementary constraints that are difficult to obtain from any single observable. This raised a broader question about whether the field currently needs primarily larger, more interconnected observational datasets or improved theoretical models. While larger and more homogeneous samples will provide stronger empirical constraints, interpreting these increasingly complex datasets will require models that can consistently connect stellar structure, magnetic activity, chemical evolution, and angular-momentum transport. The panel highlighted the importance of maintaining a close connection between observational and theoretical approaches as rotational studies move toward increasingly comprehensive stellar datasets.


\section{Summary and Future Outlook} \label{sec:summary}

The talks and posters in this session highlighted that rotation, magnetic activity, and chemical abundances are closely connected and might be more useful when treated together for better age estimation. Stellar rotation drives stellar dynamos and magnetic activity, while long-term evolution can also affect surface abundances through processes such as lithium depletion and mixing. At the same time, several open questions remain, including possible changes in rotational spin-down, the reliability of chemical clocks, and the limited overlap between asteroseismic and gyrochronological age samples. Larger and more homogeneous open-cluster samples will be crucial for resolving these issues.

The next few years should provide major advances in this area. \emph{Gaia} DR4 and the Rubin Observatory's Legacy Survey of Space and Time (LSST) will improve cluster membership and photometric constraints, and light-curve availability for rotation periods, while 4MOST and WEAVE will provide spectroscopic measurements for much larger samples of cluster stars, helping to map rotation, activity, and chemical abundances across a wider range of ages and masses. From 2027, \emph{PLATO} and \emph{Roman} are expected to extend precise asteroseismic constraints to cooler main-sequence stars, providing an important opportunity to directly connect seismic ages with gyrochronology and activity-based age relations within the same clusters. Making the most of these datasets will also require improved statistical and machine-learning approaches for membership identification and age inference.

Overall, the session demonstrated the value of bringing together different observational approaches and using open clusters as laboratories for studying stellar evolution. As increasingly large and precise open-cluster datasets become available, tracing the evolution of multiple stellar properties across clusters of different ages and masses will be key to improving stellar age estimates and placing stronger constraints on stellar structure and evolution. Continued cross-disciplinary discussions at future Cool Stars meetings will be important for connecting these efforts and turning the growing datasets into robust age-dating tools.

\section*{Acknowledgments}
The splinter session 'Open Clusters as Laboratories for Cool Star Evolution' was held on June 15, 2026, as part of Cool Stars 23 in Tokyo, Japan. We thank the SOC and LOC, as well as all the speakers, poster presenters, and panel participants for their contributions. The full program and abstracts remain available on the session website \footnote{\url{https://qmul-au.github.io/cs23-splinter-open-clusters/}}. DC and EG gratefully acknowledge support from UK Research and Innovation (UKRI) under the UK government’s Horizon Europe funding guarantee for an ERC starting grant (grant number EP/Z000890/1). DJF acknowledges support from Long term structural funding - Methusalem funding by the Flemish Government, project SOUL: Stellar evolution in full glory, grant METH/24/012, at KU Leuven. AH gratefully acknowledges support from UK Research and Innovation (UKRI) via a training grant awarded by the Science and Technology Facilities Council (STFC) (grant number ST/X50869X/1).

\bibliographystyle{cs23proc}
\bibliography{example.bib}

\end{document}